\documentclass[twocolumn,showpacs,amsmath,amssymb,%
nofootinbib]{revtex4}

\usepackage{graphicx}
\usepackage{dcolumn}
\usepackage{bm}
\usepackage{subfigure}
\usepackage{ifthen}
\begin{document}

\title{Description of nuclear octupole and quadrupole deformation
close to the axial symmetry:
Octupole vibrations in the X(5) nuclei $^{150}$Nd  and $^{152}$Sm}

\author{P.G. Bizzeti}
\email{bizzeti@fi.infn.it}
\author{A.M. Bizzeti--Sona}
\affiliation{Dipartimento di Fisica, Universit\`a di Firenze\\
I.N.F.N., Sezione di Firenze\\
Via G. Sansone 1, 50019 Sesto Fiorentino (Firenze), Italy}
\date{\today}

\begin{abstract}
The model, introduced in a previous paper, for the description
 of the octupole and quadrupole degrees of
freedom in conditions close to the axial symmetry, is used to
describe the negative-parity band based on the first octupole vibrational
state in nuclei close to the critical point of the U(5) to SU(3)
phase transition. The situation of $^{150}$Nd and $^{152}$Sm is discussed
in detail. The positive parity levels of these nuclei, and also the in-band
E2 transitions, are reasonably accounted for by the X(5) model. 
With simple assumptions on the nature of the octupole vibrations, 
it is possible
to describe, with comparable accuracy, also the negative parity sector,
without changing the description of the positive-parity part. 
\end{abstract}


\maketitle

\section{\label{S:1}Introduction}
In a previous paper~\cite{bizzeti04} (henceforth referred to as I) 
a simple model has been introduced to describe the phase transitions
in nuclear shape involving the octupole mode. This model,
valid for nuclear shapes close to the axial symmetry, has been used
to describe transitional nuclei in the Radium -- Thorium region.
The phase transition between octupole vibration and axial octupole deformation,
in nuclei which already possess a stable (axial) quadrupole deformation has
been investigated, and the model has been found to account for the properties 
of Thorium isotopes $^{226,228}$Th. In a second paper~\cite{bizzeti08}
the analysis has been extended to the case of phase transitions involving
at the same time the axial quadrupole and octupole degrees of freedom
(from harmonic vibrations around a spherical shape to a permanent, reflection
asymmetric, axial deformation). In particular, the X(5)--like nuclei
 $^{224}$Ra and $^{224}$Th were found to correspond well to this situation,
and their $K=0$ bands agree well with the model predictions, as well in the
negative parity sector as in the positive parity one~\cite{bizzeti08}. 

\begin{figure}[t]
\includegraphics[width=61mm,clip,bb=170 147  522 472]
{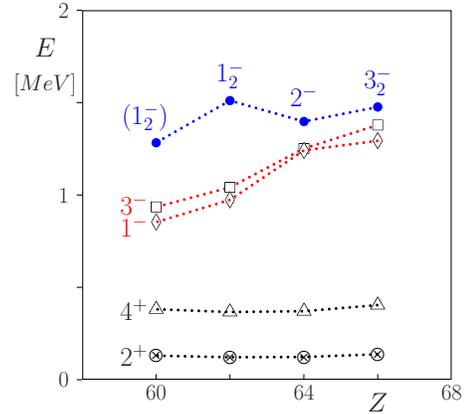}
\caption{(color on line) Excitation energies of the lowest levels
$1^-,\ 2^+,\ 3^-$ and $4^+$ in the $N=90$ isotones 
$^{150}_{\hspace*{1.2mm}60}$Nd,
$^{152}_{\hspace*{1.2mm}62}$Sm, $^{154}_{\hspace*{1.2mm}64}$Gd 
and $^{156}_{\hspace*{1.2mm}66}$Dy. The lowest negative-parity levels
not belonging to the $K^\pi=0^-$ sequence are shown as full dots.}
\label{N90}
\end{figure}

\begin{figure*}[!t]
\includegraphics[height=167mm,clip,bb=65 170 530 774]
{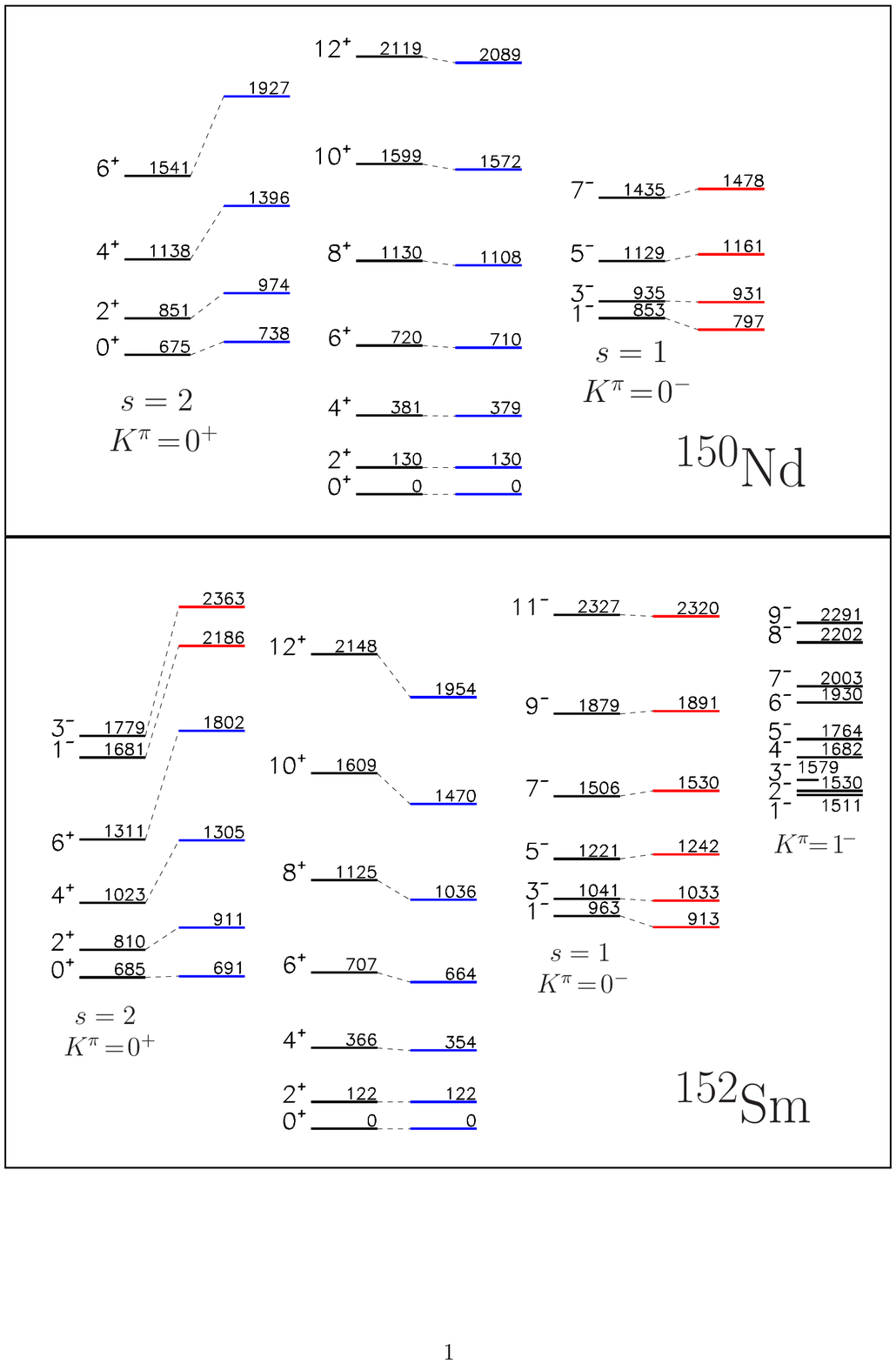}
\caption{(color on line) Partial level schemes of $^{150}$Nd and $^{152}$Sm.
For each band, the experimental values (left) are compared with the model
predictions (right) with constant $\Delta_k$:$\Delta_0=0$ (X(5) model) 
for even $J$ and parity, $\Delta_1=15$ 
or $\Delta_1=20$ for states with odd $J$ and parity of $^{150}$Nd or
 $^{152}$Sm, respectively (see Section~\ref{S:3}). 
Theoretical values of the excitation energies (in keV) are normalized to
that of the $2^+_1$ level.
Negative-parity  levels belonging to the $s=2$ sequence of $^{152}$Sm 
are shown in the same column as the positive-parity ones.
Higher lying negative parity states of $^{152}$Sm, apparently belonging to a 
$K^\pi=1^-$ band, are also shown at the right-hand margin.}
\label{F:1}
\end{figure*}

In the present part of our work, we are going to consider the effect of 
axial octupole vibrations of small amplitude, around
a reflection symmetric shape, in nuclei at the critical point between
spherical shape and axial quadrupole deformation, corresponding to the
quasi-symmetry X(5)~\cite{iachello01}.
This is apparently the
case for the two early examples of critical--point nuclei,
$^{152}$Sm~\cite{casten01} and $^{150}$Nd~\cite{kruecken02}.
In these two nuclides, the positive parity sector of the 
level scheme is well described by the X(5) 
model, while a $1^-$ state with excitation energy significantly
larger than that of the lowest $2^+$ state (see Figs.~\ref{N90},\ref{F:1}) 
can be interpreted as the lowest state
of octupole excitation. A $\Delta J=2$ band is built over this $1^-$ state,
and the absence of even-$J$ states shows that a quantum number $K=0$ can
be attributed to it, indicating an axially symmetric character for the
octupole excitation. As $K\neq 0$ bands do not appear at similar excitation
energy, this situation should be suitable to be described by the model
introduced in I.

Actually, nuclear systematics~\cite{clark03} and several
self-consistent calculations (see, {\it e.g.},
\cite{li09,robiedo08,meng05}) indicate that such a shape phase transition 
can take place in a wider region around  $^{152}$Sm. In fact,
level schemes very close to that of the X(5)
model are found also in the heavier $N=90$ nuclei $^{154}$Gd and
$^{156}$Dy (but only in $^{154}$Gd the relative B(E2) values 
follow the X(5) predictions)~\cite{tonev04,moeller06}. 
However, in these nuclei the excitation energy of the
lowest $1^-$ level increases, and approaches that of the lowest 
negative-parity level not belonging to the $K^\pi=0^-$ sequence
(Fig.~\ref{N90}):
therefore, they cannot be treated in the frame of our model,
which is intended for nuclear shapes having an approximate
(not necessarily exact) axial symmetry~\cite{bizzeti04}.

As already noted in~\cite{bizzeti02}, in the critical-point
nuclei $^{150}$Nd and  $^{152}$Sm
the level spacings in the $K^\pi=0^-$
band do not follow the same pattern of those of the ground-state band,
as it is (approximately) 
the case for the bands based on one or two phonons of $\gamma$
vibration. 
Therefore, the lowest  negative parity band cannot be explained
assuming an approximate decoupling of axial octupole and quadrupole
modes, based on the same intrinsic configuration.
Instead, we can assume that also in the present case,  
 as in the case of light Thorium isotopes
\cite{bizzeti04,bizzeti08},  an intimate connection exists between
the axial quadrupole and octupole modes, in spite of the fact that the 
latter is characterized by a much larger excitation energy.
As we shall see, a simple model based on this assumption is able to 
reproduce the negative parity band with an accuracy comparable to
that attained by the X(5) model in explaining the positive parity
states.

For convenience of the reader, the main results of our model are summarized
in the next Section~\ref{S:2}. In Sections~\ref{S:3} and \ref{S:4},
  the model is specialized to the
present case. Finally, in Section~\ref{S:5}, empirical data concerning
$^{152}$Sm and $^{150}$Nd are discussed on the basis of this theoretical model.

Preliminary results of this work have been presented at the XIV International
Workshop on Nuclear Theory (Rila, Bulgaria, 2005)~\cite{bizzetirila}
 and in the 2006 Predeal
Summer School in Nuclear Physics~\cite{bizzetipredeal}.

\section{\label{S:2}The bases of the model}
Our model is an extension of Bohr's hydrodynamic model~\cite{bohr52}, 
to include also axial octupole deformations. 
Non-axial deformations (of quadrupole or octupole character)
 are only considered up to the first order in the
corresponding parameters, as well as the variables describing the possible
misalignment of the tensors of inertia corresponding to octupole deformation or
to quadrupole deformation alone. 

Following the conventions introduced in I, we choose as the intrinsic 
reference frame
the one referred to the principal axes of the overall tensor of inertia,
resulting from the combined effects of quadrupole and octupole deformation.
Moreover, our dynamical variables
 $a^{(\lambda )}_{\mu}$ ($\lambda=2,3;\ \mu
=-\lambda ... \lambda$), describing the nuclear deformations of 
different orders, 
are considered to include the corresponding inertial parameter,
so that our variable $a_\mu^{(\lambda)}$ corresponds to 
 $\sqrt{B_\lambda}\ a^{(\lambda)}_\mu$ in the original notations of Bohr.
 The set of deformation variables
 $a^{(\lambda )}_{\mu}$ is parametrized in terms of new dynamical variables, 
\begin{eqnarray}
a^{(2)}_0 &=& \beta_2 \cos \gamma_2 \approx \beta_2\\*
a^{(2)}_1 &=& -  \frac{\sqrt{2}\ \beta_3}{\sqrt{\beta_2^2+2 
\beta_3^2}}\ v\ (\sin \varphi + i \cos \varphi )\nonumber \\*
a^{(2)}_2 &=& \sqrt{1/2}\ \beta_2\ \sin \gamma_2 - i 
\frac{\sqrt{5}\ \beta_3}{\sqrt{\beta_2^2+2 \beta_3^2}}\ u \sin \chi  
\nonumber \\*
a^{(3)}_0 &=& \beta_3 \ \cos \gamma_3 \approx \beta_3\nonumber \\*
a^{(3)}_1 &=&\frac{\sqrt{5}\ \beta_2}{\sqrt{\beta_2^2+2 \beta_3^2}} 
\ v\  (\sin \varphi + i \cos \varphi ) \nonumber \\*
a^{(3)}_2 &=& \sqrt{1/2}\ \beta_3\ \sin \gamma _3 +i
\frac{\beta_2}{\sqrt{\beta_2^2+2 \beta_3^2}}\  u \sin \chi \nonumber \\*
a^{(3)}_3 &=& w \sin \vartheta \left[ \cos \gamma_3 + 
(\sqrt{15}/2)\ \sin \gamma_3
\right]\nonumber \\*
  &+& i\ w \cos \vartheta \left[ \cos \gamma_3 - 
(\sqrt{15}/2)\ \sin \gamma_3 \right] \nonumber \\*
&\approx& w\ (\sin \vartheta +i \cos \vartheta ) \nonumber   
\label{E:2.2}
\end{eqnarray}
In the above expressions, non-axial degrees of freedom are taken into account
only up to the first order in the corresponding 
amplitudes\footnote{The simultaneous treatment of axial and 
non-axial octupole modes is
considered, {\it e.g.}, in \cite{raduta06,raduta06b} and in \cite{eisen1}.}.
Finally, the variables $\gamma_2$ and $\gamma_3$ are
expressed as
\begin{eqnarray}
\gamma_2 &=& \frac{\sqrt{10}\ \beta_3}{\beta_2
\sqrt{\beta_2^2+5\beta_3^2}}\ u\ \cos \chi 
+\frac{f(\beta_2,\beta_3)}{\sqrt{\beta_2^2+5\beta_3^2}}\ u_0 \\*
\gamma_3&=&-\frac{\sqrt{2}\ \beta_2}{\beta_3
\sqrt{\beta_2^2+5\beta_3^2}}\ u\ \cos \chi 
+\frac{\sqrt{5}\ f(\beta_2,\beta_3)}{\sqrt{\beta_2^2+5\beta_3^2}}\ u_0
\nonumber 
\label{E:2.3}
\end{eqnarray}
With this choice,
the tensor of inertia turns out to be diagonal 
{\em up to the first order in the small
variables $u_0,\ u,\ v,\ w, \vartheta,\ \varphi$, and $\chi$}. 
Up to this point, the form of the function 
$f(\beta_2, \beta_3)$ is left completely free. In \cite{bizzeti08}
a possible choice for this function is introduced,
namely $f(\beta_2,\beta_3)=\sqrt{(\beta_2^2+\beta_3^2)(\beta_2^2+2\beta_3^2)
/(\beta_2^2+5\beta_3^2)}$, to obtain the
proper basis for the description of a critical point in both
the (axial) quadrupole and octupole degrees of freedom. 
With this choice, the determinant of the matrix of inertia takes the form
\begin{eqnarray}
G\propto \frac{(\beta_2^2+\beta_3^2)^2\ (\beta_2^2+2\beta_3^2)^4}{
(\beta_2^2+5\beta_3^2)^2}
\ u_0^2 v^2 u^2 w^2
\end{eqnarray}
and the differential equation in the variables
$\beta_2$, $\beta_3$ that is obtained with the Pauli procedure~\cite{pauli} of
quantization (assuming an approximate decoupling from the part
involving all other dynamical variables) reduces to the Bohr equation 
at the limit $\beta_3\to 0$. One obtains
\begin{eqnarray}
\label{E:4}
&&\hspace*{-15mm}
\frac{1}{g}\left\{ \frac{\partial}{\partial \beta_2 } 
\left[g\frac{\partial \Psi}{\partial \beta_2 } \right] 
+\frac{\partial }{\partial \beta_3 } 
\left[g\frac{\partial \Psi}{\partial \beta_3 } \right] \right\}\\*
&+&\left\{ \epsilon - V + \frac{A_J}{\beta_2^2 +2\beta_3^2 } \right\}
\Psi(\beta_2 ,\beta_3)=0\nonumber
\end{eqnarray}
where $g\propto G^{1/2},\ V=V(\beta_2 ,\beta_3)$ and $A_J= J(J+1)/3$.
Due to the time reversal invariance, the potential must be 
even with respect to $\beta_3$, 
$V(\beta_2,-\beta_3)=V(\beta_2,\beta_3)$, while the
wavefunction $\Psi$ must be even in $\beta_3$ for the
states of even parity and $J$ and odd for states
of odd parity and $J$.

It is convenient to eliminate from Eq.~\ref{E:4} the first derivative terms, 
with the substitution
\begin{eqnarray}
\Psi = \Psi_0 / g^{1/2} .
\end{eqnarray}
The differential equation for $\Psi_0(\beta_2, \beta_3)$ is
\begin{eqnarray}
\label{E:7}
&&\hspace*{-10mm}
\frac{\partial ^2 \Psi_0}{\partial \beta_2^2} + 
\ \frac{\partial ^2 \Psi_0}{\partial \beta_3^2}\\*
&+&\left\{ \epsilon - V + V_g - \frac{A_J}{\beta_2^2 +2 \beta_3^2} \right\}
\Psi_0(\beta_2 ,\beta_3)=0\nonumber
\end{eqnarray}
with
\begin{eqnarray}
V_g&=&\frac{1}{4g^2}\left[ \left( \frac{\partial g}{\partial 
\beta_2} \right) ^2 + \left( \frac{\partial g}{\partial \beta_3} \right) ^2 
\right]-\frac{1}{2g}\left[ \frac{\partial^2 g}{\partial 
\beta_2^2} +\frac{\partial^2 g}{\partial \beta_3^2} \right]\nonumber\\*
&=&-\frac{2(\beta_2^6+37\beta_2^4\beta_3^2+
107\beta_2^2\beta_3^4+95\beta_3^6)}{(\beta_2^2+5\beta_3^2)^2
(\beta_2^4+3\beta_2^2\beta_3^2+2\beta_3^4)}
\label{E:8}
\end{eqnarray}
In the
situation considered in \cite{bizzeti08}, a flat potential extends in a rather
wide region in both directions of $\beta_2$ an $\beta_3$. 
The two-dimensional Schroedinger equation cannot be solved analytically,
but rather accurate solutions can be obtained by numerical evaluation.
We refer to \cite{bizzeti08} for more details and for the comparison
with experimental data.

In the present case, we want to consider, instead, the case where
the octupole deformation parameter $\beta_3$ is constrained to
remain always a small {\em fraction} of the quadrupole deformation
$\beta_2$. As we shall see, introducing a few simplifying assumptions,
we obtain results that can be expressed in a close form, as in the
case of the standard X(5) model.

\section{\label{S:3} Details of the model and results}
There is a deep qualitative difference between the structure of
the $K=0$, alternate parity bands of $^{152}$Sm, 
$^{150}$Nd and those of the transitional Ra  and Th isotopes 
(see Figs.7 and 9 of \cite{bizzeti08}). 
In the latter, the lowest $1^-$ level
lies between the $2^+$and the $4^+$ states, and starting
from the $5^-$ the positive- and negative-parity levels
follow each other in the increasing order of $J$.
Instead, in $^{152}$Sm and $^{150}$Nd (Fig.~\ref{F:1}),
 the negative parity
levels of angular momentum $J$ are higher than the positive-parity
ones of angular momentum $J+1$ at least up to $J=14$,
and the first $1^-$ is found between the lowest $6^+$ 
and $8^+$.
As the mean square of the deformation parameter for a particular mode of
collective excitation is inversely proportional to the energy of the
first excitation of this mode, we can conclude that, in the present
cases, the octupole deformation is much smaller than the quadrupole one.
We will assume that $\beta_3$ remains confined to a rather small
{\em fraction} of the quadrupole amplitude $\beta_2$ also at the
largest values of $J$.
As in~\cite{bonatsos05,bizzetirila}, 
we express both (axial) deformation amplitudes in terms
of two new variables, $\beta$ and $\delta$, which will be considered as 
the independent variables in the following discussion:
\begin{eqnarray}
\beta_2=\beta \cos \delta\\*
\beta_3=\beta \sin \delta \nonumber
\label{E:3}
\end{eqnarray}

Now, the Eq.~\ref{E:7} can be easily rewritten in terms of the new
variables to obtain
\begin{eqnarray}
\label{E:9}
&&\hspace*{-10mm}
\frac{\partial ^2 \Psi_0}{\partial \beta^2} + 
\frac{1}{\beta}\ \frac{\partial \Psi_0}{\partial \beta} + 
\frac{1}{\beta^2}
\ \frac{\partial ^2 \Psi_0}{\partial \delta^2}\\*
&+&\left\{ \epsilon - V + V_g - \frac{A_J}{\beta^2 (1+\sin^2 \delta)} \right\}
\Psi_0(\beta ,\delta)=0\nonumber
\end{eqnarray}
where
\begin{eqnarray}
V_g(\beta ,\delta)&=&\frac{1}{\beta^2}\left[ -2+U_g(\delta)\right]\\*
U_g(\delta)&=&-\frac{\sin^2 \delta}{1+\sin^2 \delta}
\ \frac{50+24\sin^2 \delta
+16\sin^4 \delta}{\left( 1+4\sin^2 \delta\right)^2} \nonumber
\label{E:9b}
\end{eqnarray}
The Eq.~\ref{E:9} results to be separable if the potential $V(\beta , \delta)$
takes the form
\begin{eqnarray}
V(\beta ,\delta)=V_\beta(\beta) + \frac{1}{\beta^2}U_\delta(\delta)
\end{eqnarray}
We can observe that the factor $1/\beta^2$ in the $\delta$-dependent part
of the potential becomes irrelevant if $U_\delta (\delta )$ is zero in the 
interval $-\delta_0 < \delta < \delta_0$ and $+\infty$ outside this interval
(as for a typical critical-point potential). With the above choice for 
the potential, one can put $\Psi_0=\beta^{-1/2}\psi(\beta)\phi(\delta)$ 
to obtain the independent differential equations  
\begin{eqnarray}
\label{E:12}
&&\hspace*{-6mm}\frac{{\rm d}^2 \phi_k(\delta)}{{\rm d} \delta^2}
+\left[A^\prime_k(J) - \tilde{U}_\delta(\delta) +
\frac{A_J \sin ^2 \delta}{1+\sin ^2 \delta} \right] \ \phi_k(\delta)=0\\*
&&\hspace*{-6mm}\frac{{\rm d}^2 \psi_k(\beta)}{{\rm d} \beta^2}
+\left[\epsilon_k-\tilde{V}_\beta (\beta)-\frac{A_J+2+\Delta_k(J)}{\beta^2}
\right] \ \psi_k(\beta)=0 \nonumber
\end{eqnarray}
where $\tilde{U}_\delta=U_\delta - U_g$,
while $\Delta_k(J)=A^\prime_k(J)-A^\prime_0(0)$ and
$\tilde{V}_\beta(\beta)=V_\beta(\beta)+[A^\prime_0(0)-1/4]/\beta^2$.
 The index $k=0$ corresponds
to even parity and $J$, $k=1$ to odd parity and $J$.

The equation in $\delta$ contains a weak dependence on $J$ in the
term $A_J$. We shall see that the 
resulting dependence on $J$ of the separation constant $A^\prime_k$ is
actually negligible, at least for not too large values of $J$.
If we neglect it, 
$\Delta_0=0$ and $\Delta_1\equiv A^\prime_1-A^\prime_0$
is a constant. We will
consider it as an adjustable parameter.
We can expect that this approximation remains valid also for different
choices of the potential for $\delta$, as long as Eq.~\ref{E:9} remains
at least approximately separable.

We must now do some assumptions on the $\beta$ dependent part of 
the potential. We know that the positive--parity
part of the level scheme is in good agreement with the X(5) predictions.
Therefore, the potential term $\tilde{V}_\beta$ ({\em not the potential $V_\beta$!})
will be taken constant in the interval $0<\beta \leq \beta_w$ and 
$=+\infty$ outside, as in the X(5) approximation.

This choice for the potential deserves some more comments. 
Actually $\tilde{V}_\beta$ contains, in addition to the original
$V_\beta$, the zero-point energy of the octupole vibrations, which
turns out to depend on the value of $\beta$. We must observe, however,
that the ``model potential'' which determines the properties of the
motion in the $\beta$ degree of freedom certainly include the zero-point
energies of all ignored degrees of freedom of the system, first-of-all
those related to the single-nucleon ones, which certainly depend
on the deformation parameters. The evolution of this 
 effective potential determines the pattern of the phase transition,
and in particular the position of the critical point.

\begin{center}
\begin{figure}[t]
\includegraphics[height=75mm,angle=90,clip,bb=65 75 575 722]
{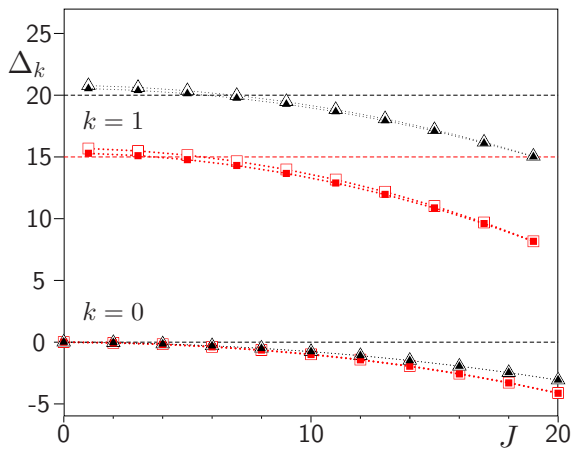}
\caption{(color on line) Values of $\Delta_k(J)=A^\prime_k(J)-A^\prime_0(0)$,
as a function of $J$, with different assumptions on the potential-like
term $U_\delta$ (or $\tilde{U}_\delta$). The index $k=0$ corresponds
to even parity and $J$, $k=1$ to odd parity and $J$.
Open symbols: $U_\delta =0$ for $-\delta_w<\delta <\delta_w$
and $=+\infty$ outside, with $\delta_w=0.73$ (squares, for $^{150}$Nd) or 
$\delta_w=0.62$ (triangles, for $^{152}$Sm).
Full symbols: for $\tilde{U}_\delta=0$ (instead of
$U_\delta=0$), and $\delta_w=0.695$ or $\delta_w=0.60$, respectively. 
The horizontal dashed lines correspond to the constant values
used to approximate $\Delta_k(J)$ for $^{152}$Sm and $^{150}$Nd 
($\Delta_1=20$ and $\Delta_1=15$, respectively).}
\label{F:2}
\end{figure}
\end{center}

It remains now to  solve the $\beta$ dependent equation (second line
 of Eqs. \ref{E:12}) with $\tilde{V}_\beta=0$
and with the boundary conditions $\psi(0)=\psi(\beta_w)=0$.
Neglecting the $J$ dependence of $\Delta_k$, and with the
new substitutions $\beta =z / \sqrt{\epsilon_k}$, 
\ $\psi(\beta)=\sqrt{z}\ \xi_\nu(z)$, we can transform this equation
into the classical Bessel equation of order $\nu=\nu_k\equiv \sqrt{A_J+9/4+\Delta_k}$:
\begin{eqnarray}
\frac{{\rm d}^2}{{\rm d}z^2}\xi_\nu(z)+
\frac{1}{z}\frac{\rm d}{{\rm d}z}\xi_\nu(z)+\left[ 1-\frac{\nu^2}{z^2}
\right] \xi_\nu(z)=0
\label{E:13}
\end{eqnarray}

\noindent
The boundary condition at the upper border is satisfied if
$z_s(\nu)\equiv \beta_w \sqrt{\epsilon_s}$ is the $s^{th}$ zero of the Bessel
function $J_{\nu}(z)$, and therefore the eigenvalues of $\epsilon$ 
are given by
\begin{eqnarray}
\epsilon_{k,s}= \left[z_s(\nu_k)/\beta_w \right]^2
\end{eqnarray}
For $k=0$ (even parity and $J$), $\Delta_0=0$ and we obtain again, as expected,
the X(5) level scheme. For $k=1$, we obtain the level sequence of the odd
parity, odd--$J$ part of the band. The comparison with experimental level schemes
of $^{152}$Sm and $^{150}$Nd  (Fig~\ref{F:1}) shows that a satisfactory
agreement can be obtained assuming a constant $\Delta_1$ equal to $20$ for
$^{152}$Sm and to $15$ for $^{150}$Nd.

To check the effect of the $J$ dependent term in the first of Eqs.~\ref{E:12},
it is necessary to assume a definite form for the potential $U_{\delta}$ (or,
if we prefer, for $\tilde{U}_{\delta}$). As both $A_1^\prime$ 
and $A_0^\prime$ become a
function of $J$, we must be prepared to find some differences between the
level scheme resulting from the present model and the one of X(5), also for
the ground-state band. It is matter to see how large these differences are.
\begin{center}
\begin{figure}[b]

\includegraphics[height=78mm,angle=90,clip,bb=130 70 575 761]
{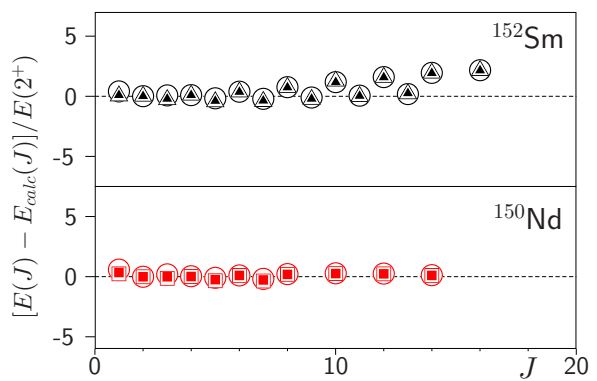}
\caption{(color online) Differences between the experimental energy of
levels in the $K=0^\pm$ bands of $^{150}$Nd and $^{152}$Sm and the 
calculated values obtained with different assumptions, divided by the
energy of the first excited state. Open circles
correspond to the assumption of constant $\Delta_k$: $\Delta_0=0$,
 {\it i.e.} X(5) model, for even parity and $J$ and $\Delta_1=15$
(or $=20$) for odd parity and $J$ in $^{150}$Nd (or $^{152}$Sm).
Other symbols have the same meaning as in Fig.~\ref{F:2}.
}
\label{F:3}
\end{figure}
\end{center}
In Fig.~\ref{F:2}, the dependence of $\Delta_k$ on the angular momentum is 
depicted for situations close to the experimental ones of $^{152}$Sm
and $^{150}$Nd. In Fig.~\ref{F:3}, 
 the differences between the experimental level energies (in units of
$E(2^+)$) and those deduced with these $J$-dependent values of $\Delta_k$ 
or corresponding to constant $\Delta_k$ are also reported. The differences
between the different calculated values are
remarkably small, and usually smaller than their
deviations from the experimental ones. This result is not surprising:
in fact, at small values of $J$ the differences between the 
values of $\Delta_k$
resulting from the different assumptions are quite small, while at large
values of $J$ they result to be almost insignificant, in comparison to 
the large values of $A_J$. The difference between the different 
calculated values
in the $s=2$ sector (not shown in the figure) are also very small 
($<0.1$ for positive parity states and $<0.7$ for negative parity ones).

In most cases, the experimental
values are satisfactorily reproduced. The larger deviations
observed in $^{152}$Sm are limited to the even $J^\pi$ values (X(5) states)
above $8^+$,
while -- surprisingly enough -- the agreement is much better for the
odd $J$ values.

It remains to verify whether the average value of $\beta_3^2$ 
are really small enough in comparison to the average $\beta^2$,
for the values of $\Delta_1$ which reproduce the experimental data.
Actually, for a square-well potential giving $\Delta_1=20\ (15)$, 
the value of $< \sin^2 \delta > $
is about $0.05\ (0.09)$ for the positive parity part and
$0.10\ (0.14)$ for the negative-parity part of the $K=0$ band. 
\begin{table}
\caption{\label{T:e1}Measured values \cite{garrett09,nndc} 
 of B(E1) in $^{152}$Sm, compared with calculated ones
for constant $\Delta_1=20$,
normalized to the $1^-\to 0^+$ transition. 
}
\begin{ruledtabular}
\vspace{0.4cm}
\begin{tabular}{ccccc}
\vspace{-0.20cm}
& & & & \\
Transition & E$_i$& E$_{\gamma}$
&\multicolumn{2}{c}{B(E1) [$10^{-3}$ W.u]}\\
$s_i,J^\pi_i\to s_f,J^\pi_f$& [keV] & [keV] & Experimental & Calculated \\
\vspace{-0.20cm}
& & & & \\
\hline
\vspace{-0.20cm}
& & & &  \\
$1,1^- \to 1,0^+ $& 963 & 963 & 4.2\ (4) & 4.2 (norm.)\\
$1,1^- \to 1,2^+ $& 963 & 841 & 7.7\ (7) & 9.8 \\
$1,3^- \to 1,2^+ $& 1041 &919 & 8.0\ (16)& 6.3 \\
$1,3^- \to 1,4^+ $& 1041 & 675 & 8.4\ (4) & 10.0$\phantom{1}$ \\
$1,5^- \to 1,4^+ $& 1222 & 855 & 4.1\ (9) & 8.0 \\
$1,5^- \to 1,6^+ $& 1222 & 675 & 5.0\ (12) & 11.1$\phantom{1}$\\
& & & & \\
$2,1^- \to 2,0^+ $& 1681 & 996 & 2.1\ (2) & 1.93\\
$2,1^- \to 2,2^+ $& 1681 & 870 & 5.4\ (5) & 5.21\\
$2,3^- \to 2,2^+ $& 1779 & 969 & 2.4\ (5) & 3.14\\
$2,3^- \to 2,4^+ $& 1779 & 756 & 3.9\ (8) & 5.91\\
& & & &\\
$1,1^- \to 2,0^+ $& 963  & 279 & (weak)     & 2.70\\
$1,1^- \to 2,2^+ $& 963  & 153 & 0.13\ (4)  & 4.47\\
& & & &\\
$2,1^- \to 1,0^+ $& 1681 & 1681 & 0.041\ (6) & 0.001\\
$2,1^- \to 1,2^+ $& 1681 & 1559& 0.076\ (9) & 0.021\\
$2,3^- \to 1,2^+ $& 1779 & 1657 & 0.019\ (4) & 0.001\\
$2,3^- \to 1,4^+ $& 1779 & 1413 & 0.019\ (4) & 0.109
\end{tabular}
\end{ruledtabular}
\end{table}

\section{\label{S:4}The transition amplitudes}

In the frame of the X(5) model, the E2 transition operator is defined as
$\mathcal{M}(E2) \propto \beta^2 Y^{(2)}$, and its matrix elements 
can be easily 
evaluated by means of 
standard tensor algebra~\cite{bizzeti08}, as well for the negative parity as
for the positive parity sector. Unfortunately, in-band E2 
transitions between negative parity states cannot compete with the predominant
E1 decay. Three inter-band transitions between equal-parity states
have been reported \cite{garrett09} 
but two of them can be mixed M1-E2, with unknown mixing ratios.
Therefore, in order to check the validity of the present 
extension of the X(5) model, it is necessary to investigate the E1 
transitions. 

In the limit of the original Bohr model ({\it i.e.}, assuming
a constant charge density of the nuclear matter) all E1 transitions are
strictly forbidden. If one assumes that the nuclear matter possesses, to some
extent, an {\em electric polarizability}, the E1 operator takes the 
form~\cite{bohr57,bohr58}
$\mathcal{M}(E1) \propto \beta_2 \beta_3 Y^{(1)}$
that we actually used in our previous works \cite{bizzeti04,bizzeti08}.
Relative values of the E1 transition amplitudes have been evaluated in the 
frame of the present model, using  the wavefunctions corresponding to the 
value  $\Delta_1=20$ appropriate for $^{152}$Sm. 
In Table~\ref{T:e1}, calculated values of B(E1) are compared with available
experimental data~\cite{garrett09,nndc}.
\begin{table}[!h]
\caption{\label{T:e2}Measured \cite{casten01,garrett09} and calculated 
values of B(E2) in $^{152}$Sm, normalized to the $2^+\to 0^+$ transition. 
New results, 
concerning the negative parity sector \cite{garrett09}, are shown in bold.}
\begin{ruledtabular}
\vspace{0.4cm}
\begin{tabular}{ccccc}
\vspace{-0.20cm}
& & & & \\
Transition & E$_i$ & E$_{\gamma}$ 
&\multicolumn{2}{c}{B(E2) [W.u]}\\
$s_i,J^\pi_i\to s_f,J^\pi_f$& [keV] & [keV] & Experimental & Calculated \\
\vspace{-0.20cm}
& & & & \\
\hline
\vspace{-0.20cm}
& & & &  \\
$1,2^+ \to 1,0^+ $& 122 & 122 & 144\ (4) & 144 (norm.)\\
$1,4^+ \to 1,2^+ $& 366 & 244 & 209\ (7) & 230 \\
$1,6^+ \to 1,4^+ $& 707 & 341 & 245\ (16)& 285 \\
$1,8^+ \to 1,6^+ $& 1125 & 418 & 285\ (4) & 328 \\
$1,10^+\to 1,8^+ $& 1609 & 484 & 320\ (9) & 361 \\
& & & & \\
$2,2^+ \to 2,0^+ $& 810 & 125 & 111\ (2) & 114\\
$2,4^+ \to 2,2^+ $& 1023 & 213 & 204\ (5) & 173\\
& & & &\\
$2,0^+ \to 1,2^+ $& 685 & 573 & 33\ (6) & 83\\
$2,2^+ \to 1,0^+ $& 810 & 810 & 1\ (9) & 3\\  
$2,2^+ \to 1,2^+ $& 810 & 688 & 3\ (4) & 11\\ 
$2,2^+ \to 1,4^+ $& 810 & 444 & 19\ (4) & 49\\ 
$2,4^+ \to 1,2^+ $& 1023 & 901 & 1\ (4) & 1\\  
$2,4^+ \to 1,4^+ $& 1023 & 657 & 5\ (4) & 8\\  
$2,4^+ \to 1,6^+ $& 1023 & 316 & 4\ (4) & 37\\ 
& & & &\\
\bf$ 2,1^- \to 1,3^- $&\bf  1681 &\bf 640 &\bf 8.4 \ (17) &\bf 19\\
\bf $2,1^- \to 1,1^- $&\bf  1681 &\bf  268 &\bf  $\bm{\leq }$11\ (2) &\bf  10\\
\bf $2,3^- \to 1,3^- $&\bf  1779 &\bf  738 &\bf  $\bm{\leq }$46\ (9) &\bf  6\\
\end{tabular}
\end{ruledtabular}
\end{table}

\section{\label{S:5}Comparison with experimental data}
In Fig.~\ref{F:1}, the calculated level energies are compared with the
experimental ones
for $^{150}$Nd and $^{152}$Sm. As for the positive parity
states, our results obviously coincide with the standard X(5)
model\footnote{A few small deviations from the
published values are presumably due to different rounding errors.}.
The negative parity states are evaluated in the present model, using only one
additional parameter ($\Delta_1$). For the $s=1$ negative parity band, the 
observed agreement (Fig.~\ref{F:3}) is comparable with (and perhaps better 
than) the one reported for
the positive parity states. In both cases, however, the first $1^-$ level
results to be somewhat  lower than the experimental one.

As for the $s=2$ part of the spectrum, it is well known that
the position of the 
$0^+_2$  predicted by the X(5) model~\cite{iachello01} is in satisfactory 
agreement with the experimental one, but the energy spacings between positive
parity levels are appreciably larger than the experimental values.
In $^{152}$Sm, we have some information also on the negative parity states
of the $s=2$ sector. The tentative identification~\cite{bizzetipredeal} 
of the $1^-$ state at 1681 keV as the lowest state of the negative parity
part of this band has been confirmed by the recent 
measurements by Garrett {\it et al.}~\cite{garrett09},
who also identified the next level, with $J^\pi=3^-$, at  1779 keV.
Both levels are much lower than those calculated in the present model:
we can conclude that this is a systematic effect for all $s=2$ levels
predicted in the frame of the X(5) model, apart from the $0^+$ band head.

\begin{center}
\begin{figure}[!t]
\includegraphics[width=80mm,clip,bb=80 424 496 750]
{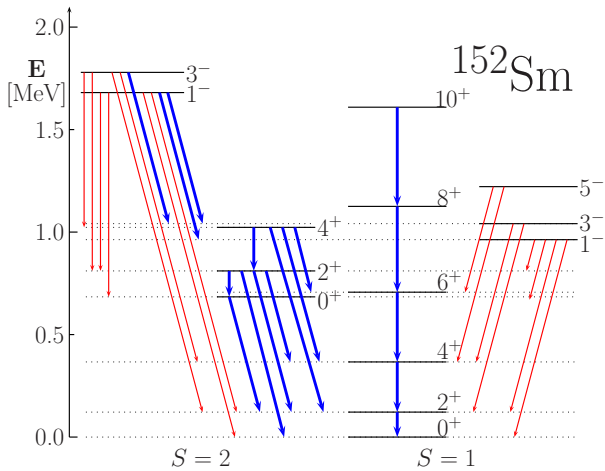}
\caption{\label{F:tr}(color on line) 
Experimentally observed transitions  
in $K=0$ bands of $^{152}$Sm, whose strengths are reported 
in \cite{casten01,garrett09,nndc}. Thinner lines: E1 transitions;
thicker ones: E2 or E2/M1.
 Intra-band transitions are shown as
vertical lines, inter-band ones as inclined lines.}
\end{figure}
\end{center}

A more sensitive test of the model, concerning the E1 electromagnetic 
transitions,
is only possible for   $^{152}$Sm 
(see Fig.~\ref{F:tr}).
In Table~\ref{T:e1}, the experimental values of E1 reduced strengths 
are compared with the calculated ones.
Since the value of $\Delta_1$ used in the present
calculation has been fixed by level energies, no additional free parameters
are used, apart from a common factor of scale which has been
normalized to the $B$(E1,$1^-\to 0^+$).
For the in-band transitions, the agreement is
satisfactory up to the $3^-$ state of both the $s=1$ and $s=2$ bands.
Experimental B(E1) values for transitions from the $s=1,5^-$ 
state result to be appreciably smaller than expected. 
A large disagreement is found for the two inter-band transitions from the
$s=1, 1^-$ to the $s=2, 0^+$ and $2^+$ levels, for which our model
predicts a relatively large value of B(E1), at variance with experimental 
results reported in the NNDC tabulation~\cite{nndc}.
For the $s=2\to s=1$ inter-band transitions, the 
model predicts very small values of $B$(E1). Experimental values
are also very small, but not in agreement with the calculated ones.
One has, however, to remember that also in the positive parity sector, a
significant disagreement is found between X(5) predictions and
experimental $B$(E2) values for inter-band transitions.

For comparison, X(5) predictions and available experimental values
for reduced E2 strengths in the positive parity sector are reported
in Table~\ref{T:e2}. Finally, in the last three rows of 
Table~\ref{T:e2}, one known value and two upper limits
for the reduced E2 strengths from $s=2$ to the $s=1$ negative parity 
levels, deduced from the new experimental results of
\cite{garrett09} are compared with the calculated values. 
The experimental B(E2) for the $s=2,1^-\to s=1,3^-$ transition results to be
less than one half of the calculated one. For the 
 $s=2,1^-\to s=1,1^-$ and  $s=2,3^-\to s=1,3^-$ transitions, 
only upper limits of B(E2) are known, due to the possible mixing of M1
multipolarity. These limits are consistent with the
calculated values.

\section{\label{S:6}Discussion}
We can conclude that the present model 
is able to satisfactorily account for the lowest octupole excitations 
of $^{150}$Nd and $^{152}$Sm.
From the comparison of level energies, also in the
negative-parity sector, both $^{150}$Nd and $^{152}$Sm (and, perhaps,
more $^{150}$Nd than $^{152}$Sm) result to be very close to the critical-point
behavior, as it is defined by X(5) and by the present model. 

As for the electromagnetic transition strengths,
sufficient data on E1 transitions are only available for $^{152}$Sm. 
The comparison of these data with our model predictions shows
a comparable degree of success and comparable limitations as for the
E2 transitions in the original X(5) model. In fact, for 
both nuclei, the E2 strengths are in satisfactory agreement 
with X(5) predictions in the case of
lower lying intra-band transitions, while for
inter-band transitions the calculated values are usually larger than
experimental ones~\cite{casten01,kruecken02}. 
This fact shows that the overlap between $s=1$ and
$s=2$ wavefunctions resulting from the square-well potential is too
large. Perhaps a slightly different potential well as,
{\it e.g.}, the one 
discussed by Caprio~\cite{caprio04}, could improve the agreement with experimental
data, also for what concern the level spacing in the $s=2$ band,
but at the expenses of having one more parameter in the model.
The failure of the model to reproduce the high-spin levels,
and the electromagnetic transitions between them, can be ascribed to
several possible reasons. First, we must note that other levels with
the same $J^\pi$ as the ones of the $s=1$ or $s=2$ bands appear above
an excitation energy of about 1.2 MeV.
Part of them also have collective character, as those of the $K^\pi=1^-$
band based on the 1511 keV level of $^{152}$Sm (see Fig.~\ref{F:1}),  
and probably mix, to some extent,
with those considered here, due to Coriolis interactions.
Moreover, the internal structure of the rotating and oscillating nucleus
can be altered by the effect of Coriolis forces in their non-inertial
reference frame, and the shape of the effective potential in the collective
coordinates could change accordingly~\cite{williams08,cejnar03,cejnar02}.

We must remind that a model like X(5) (and its present extension
to the axial octupole mode) is not intended to be able of reproducing the 
properties of an entire class of nuclei, but is more like a 
bench-mark~\cite{casten03}
saying how close a given nucleus is to the critical point of the shape
phase transition.  
It would not be surprising, therefore, if a similar degree of agreement can
be obtained in the frame of a more general model involving a much larger
number of adjustable parameters, such as the $spdf$-IBM
\cite{garrett09,babilon05} or the coherent-coupling model by 
Minkov {\it et al.}~\cite{minkov06}.

Within its obvious limits, however, the proposed model seems to be able to
reproduce (with only one more parameter)
the $K^\pi=0^-$ octupole bands of $^{150}$Nd and $^{152}$Sm
and at least the in-band E1 transitions in  $^{152}$Sm with a comparable
degree of accuracy as the original X(5) model does for the
positive-parity ones.

\bibliography{biz_bib}
\end{document}